\documentclass[a4paper,14pt]{article}
\usepackage[english]{babel}
\pdfoutput=1 

\usepackage{jheppub} 

\usepackage[T1]{fontenc} 

\usepackage[T1,T2A]{fontenc}
\usepackage[utf8x]{inputenc}

\usepackage{mathrsfs}

\usepackage{comment}

\usepackage{cases}

\usepackage{mathtools}

\newcommand{\bi}{\begin{itemize}}
    \newcommand{\ei}{\end{itemize}}
\newcommand{\bea}{\begin{eqnarray}}
    \newcommand{\eea}{\end{eqnarray}}
\newcommand{\bt}{\begin{tabular}}
    \newcommand{\et}{\end{tabular}}
\newcommand{\bc}{\begin{center}}
    \newcommand{\ec}{\end{center}}

\newcommand{\be}{\begin{equation}}
    \newcommand{\ee}{\end{equation}}
\newcommand{\ba}{\begin{array}}
    \newcommand{\ea}{\end{array}}

\def\bbox{{\,\lower0.9pt\vbox{\hrule \hbox{\vrule height 0.2 cm
                \hskip 0.2 cm \vrule height 0.2 cm}\hrule}\,}}
\newcommand{\dsl}{\pa \kern-0.5em /}

\makeatletter \@addtoreset{equation}{section} \makeatother

\def\slashchar#1{\setbox0=\hbox{$#1$}           
    \dimen0=\wd0                                 
    \setbox1=\hbox{/} \dimen1=\wd1               
    \ifdim\dimen0>\dimen1                        
    \rlap{\hbox to \dimen0{\hfil/\hfil}}      
    #1                                        
    \else                                        
    \rlap{\hbox to \dimen1{\hfil$#1$\hfil}}   
    /                                         
    \fi}

\pdfoutput=1

\title{\boldmath  Novel $\mathcal{N}=2$ higher-spin supercurrents}

\author[]{Nikita~Zaigraev}

\affiliation[]{Bogoliubov Laboratory of Theoretical Physics, JINR,\\141980 Dubna, Moscow region, Russia}
\affiliation[]{Moscow Institute of Physics and Technology,\\ 141700 Dolgoprudny, Moscow region, Russia}

\emailAdd{nikita.zaigraev@phystech.edu}

\abstract{ We study cubic interactions of $\mathcal N=2$ massless integer-spin gauge supermultiplets in harmonic superspace. We construct the complete class of abelian $(\mathbf{s},\mathbf{s_1},\mathbf{s_2})$ cubic vertices with the minimal number of space-time derivatives. Such vertices exist only for $\mathbf{s}\geq \mathbf{s_1}+\mathbf{s_2}$ and universally take the form of the gauge prepotential coupled to the conserved higher-spin supercurrent. For~$\mathbf{s_1}\neq\mathbf{s_2}$, we find the novel complex principal supercurrent, whose real and imaginary parts generate the parity-invariant and the parity-breaking interactions, respectively. The supercurrents are constructed from gauge-invariant $\mathcal N=2$ higher-spin Weyl supertensors associated with the spin-$\mathbf{s_1}$ and spin-$\mathbf{s_2}$ gauge multiplets. These supertensors are defined in terms of unconstrained higher-spin analytic prepotentials.  
We also derive the complete set of conserved component higher-spin currents associated with the $(s,s_1,s_2)$ vertices, including both traceless currents and currents with the non-vanishing trace.
\vspace{0.5cm}

\begin{flushright}
	\textit{In honor of the 70th anniversary of the Bogoliubov Laboratory of Theoretical Physics}
\end{flushright}
 }

\arxivnumber{2606.05370 }

\makeatletter
\gdef\@fpheader{}
\makeatother

\begin{document}

\maketitle
\flushbottom


\section{Introduction}

According to the Metsaev classification  of cubic vertices $(s, s_1,s_2)$  in the four-dimensional Minkowski space \cite{Metsaev:2005ar, Metsaev:2007rn}, there are two types of parity-invariant cubic vertices (for integer $s\geq s_1\geq s_2$) having either $s+s_1+s_2$ or $s+s_1-s_2$ derivatives. Among vertices with a minimal number of derivatives, the abelian ones are of particular interest due to their simple structure, which can be schematically represented as follows:
\begin{equation*}
\mathcal{L}_{\text{abelian}} \sim \Phi C C,
\end{equation*}
where $\Phi$ denotes a higher-spin field, and $C$ denotes a higher-spin Weyl-like tensor. Such vertices exist provided the following inequality on the spins holds:
$$
s \geq s_1 +s_2.
$$ 
One can construct such interactions using higher-spin currents.
Examples of such higher-spin currents were first considered for the $s_1=s_2$ case in \cite{Berends:1985xx}. Particular cases of these interactions have also been studied in \cite{Manvelyan:2009vy, Zinoviev:2010cr, Gelfond:2006be}. To the best of our knowledge, such interactions in their most general form in the Lorentz-covariant formalism have not yet been considered in the literature. Such vertices are contained within Vasiliev's theory \cite{Vasiliev:1990en, Vasiliev:1992av}, as demonstrated by the results of \cite{Gelfond:2017wrh, Misuna:2017bjb, Tatarenko:2024csa}.

The supersymmetric generalizations of higher-spin currents in the  four-dimensional Minkowski  space have been actively studied using the superfield formalism \cite{Gates:1983nr, Wess:1992cp, BK, 18}. However, most of the obtained results are dedicated to interactions with matter multiplets and the construction of the corresponding higher-spin supercurrents, see Refs. \cite{Kuzenko:2017ujh, Hutomo:2017phh, Koutrolikos:2017qkx, Buchbinder:2018wwg, Buchbinder:2018gle} for $\mathcal{N}=1$ and Refs. \cite{Buchbinder:2022kzl, Buchbinder:2022vra, Kuzenko:2023vgf,  Buchbinder:2024pjm, Kuzenko:2024vms} for $\mathcal{N}=2$ supersymmetry.
To date, interactions with gauge supermultiplets have been constructed only for the $\mathbf{s_1} = \mathbf{s_2}$ case in \cite{Buchbinder:2018wzq, Gates:2019cnl, Zaigraev:2024ryg,  Zaigraev:2026jvu}\footnote{To denote supermultiplets, we use a bold letter corresponding to the value of the highest physical spin in the multiplet. For example, $\mathcal{N}=2$ spin-$\mathbf{s}$ supermultiplet contains physical fields with spin values  $\{s,2\times (s-1/2), s-1\}$. }.

 In this paper, we construct the most general class of abelian $\mathcal{N}=2$ vertex $(\mathbf{s}, \mathbf{s_1}, \mathbf{s_2})$ with a minimal number of derivatives for the general case $\mathbf{s_1}\neq \mathbf{s_2}$. 
 In addition to the superfield construction, we perform a complete component analysis of the corresponding higher-spin currents. We derive the most general conserved higher-spin currents associated with the abelian $(s,s_1,s_2)$ vertices. Solving the current conservation equations, we obtain the complete class of complex currents, including currents with the non-vanishing trace, and identify the corresponding trivial (``fake'') interactions generated by local field redefinitions. After factorization by such vertices, a single nontrivial complex current remains in the generic case $s_1\neq s_2$, giving rise to the parity-invariant and the parity-breaking cubic interactions. We also analyze the special case $s_1=s_2$, where additional reality conditions reduce the number of independent conserved currents and leave only a single nontrivial interaction with parity $(-1)^{s}$.
 
 Our superfield construction is based on the results of the work \cite{Zaigraev:2024ryg}, where it was shown that $\mathcal{N}=2$ abelian cubic interactions can be constructed using the $\mathcal{N}=2$ principal supercurrent $\mathcal{J}_{\alpha(s-2)\dot{\alpha}(s-2)}$ satisfying rather simple equations. In this formalism, the cubic vertex is expressed via the harmonic generalization of the Mezincescu-type prepotentials and is given by:
 \begin{equation}
 	S = \int d^4xd^8\theta du\, \left(\Psi^{-(\alpha(s-2)\beta) \dot{\alpha}(s-2)} \mathcal{D}^+_{\beta}
 	+
 	\bar{\Psi}^{-\alpha(s-2)(\dot{\alpha}(s-2)\dot{\beta})} \bar{\mathcal{D}}^+_{\dot{\beta}} 
 	\right)
 	\mathcal{J}_{\alpha(s-2)\dot{\alpha}(s-2)}.
 \end{equation}	
The Mezincescu-type prepotentials provide an alternative description of the degrees of freedom for off-shell $\mathcal{N}=2$ higher-spin multiplets \cite{Zaigraev:2026jvu, Buchbinder:2021ite, Buchbinder:2022vra}. Their relation to the analytic spin-$\mathbf{s}$ prepotentials~\cite{Buchbinder:2021ite}\footnote{For the generalization of analytic prepotentials to $\mathcal{N}=1, 5D$ higher-spin supermultiplets see \cite{Buchbinder:2025yef}.}, namely,
 \begin{equation}\label{eq: analyt prep}
 	h^{++\alpha(s-1)\dot{\alpha}(s-1)},
 	\quad
 	h^{++\alpha(s-1)\dot{\alpha}(s-2)},
 	\quad
 	h^{++\alpha(s-2)\dot{\alpha}(s-1)},
 	\quad
 	h^{++\alpha(s-2)\dot{\alpha}(s-2)},
 \end{equation}
 is given by the following cross-relation:
 \begin{equation}\label{eq: diff operator}
 	\begin{split}
 		\hat{\mathcal{H}}^{++}_{(s)}
 		&=
 		h^{\alpha(s-2)\dot{\alpha}M}\partial_M \partial^{s-2}_{\alpha(s-2)\dot{\alpha}(s-2)}
 		\\&=
 		(\mathcal{D}^+)^4 \left[ \Psi^{-\alpha(s-1)\dot{\alpha}(s-2)} \mathcal{D}^{-}_{\alpha}
 		+
 		\bar{\Psi}^{-\alpha(s-2)\dot{\alpha}(s-1) } \bar{\mathcal{D}}^{-}_{\dot{\alpha}} \right] \partial^{s-2}_{\alpha(s-2)\dot{\alpha}(s-2)}.
 	\end{split}
 \end{equation}	
 Here $M = \{\alpha\dot{\alpha}, \alpha+, \dot{\alpha}+, 5 \}$ and $\partial_M := (\partial_{\alpha\dot{\alpha}}, \partial^-_\alpha, \partial^-_{\dot{\alpha}}, \partial_5)$.
 
 Note that the analytic prepotentials \eqref{eq: analyt prep} provide a natural, unconstrained description of the $\mathcal{N}=2$ off-shell higher-spin prepotentials in harmonic superspace \cite{18, Galperin:1984av}. Furthermore, they directly generalize the off-shell $\mathcal{N}=2$ supergravity prepotentials \cite{18, Galperin:1987ek, Galperin:1987em, Ivanov:2022vwc} and possess the elegant geometric interpretation; see, e.g., \cite{Buchbinder:2025ceg}. However, while analytic prepotentials are geometrically fundamental, using the harmonic prepotentials of the Mezincescu-type proves to be technically more convenient for constructing this class of vertices. 
 
 Nevertheless, the resulting cubic interactions can still be recast into an analytic form. As was shown in \cite{Zaigraev:2026jvu}, on the equations of motion, the analytic supercurrents obtained in the expansion of the principal supercurrent vanish, except for a specific set:
 \begin{equation*}
 	J^{++}_{\alpha(s-1)\dot{\alpha}(s-1)},
 	\qquad
 	J^+_{\alpha(s-1)\dot{\alpha}(s-2)},
 	\qquad
 	\bar{J}^+_{\alpha(s-2)\dot{\alpha}(s-1)},
 	\qquad
 	J^{++}_{\alpha(s-2)\dot{\alpha}(s-2)}.
 	\end{equation*}
 		This non-vanishing structure allows one to represent the cubic vertex as an integral over analytic superspace, enabling an efficient analysis of both the component structure and the corresponding higher-spin supergauge transformations.
 
 \medskip

Our paper is organised as follows: in Section \ref{sec: 2}, we present the $(3,2,1)$ vertex and the corresponding $\mathcal{N}=2$ superfield  vertex $(\mathbf{3}, \mathbf{2}, \mathbf{1})$. Using this simple example, we will illustrate the main features characteristic of the interactions under consideration. Section \ref{sec: 3} is devoted to the construction of the general spin-$s$ current, which corresponds to the abelian $(s,s_1,s_2)$ interaction. This serves as a useful guide for the subsequent superfield construction. In Section \ref{sec: 4}, we construct $\mathcal{N}=2$ complex supercurrent, which corresponds to the superfield vertex $(\mathbf{s}, \mathbf{s_1}, \mathbf{s_2})$. 
In Appendix~\ref{app}, we give the definition of the $\mathcal{N}=2$ higher-spin Weyl-like  supertensor and discuss its properties, which we use in this work.


\section{Four-derivative $(3,2,1)$ vertex and its $\mathcal{N}=2$ superfield generalization}\label{sec: 2}

We start with the example of
the $(3,2,1)$ vertex with four derivatives. Such an interaction can be constructed as
\begin{equation}\label{eq: (321) vertex}
S_{(3,2,1)} = \int d^4x\; \left( 	\Phi^{\alpha(3)\dot{\alpha}(3)}\, J_{\alpha(3)\dot{\alpha}(3)} 
	+
	\Phi^{\alpha\dot{\alpha}} T_{\alpha\dot{\alpha}}  \right),
\end{equation}	
where the spin-$3$ current and the current trace must be constructed from the gauge-invariant linearized Weyl tensor and the Faraday tensor (electromagnetic field tensor).  Moreover, the current and the current trace  must be real.  The requirement of the invariance under gauge transformations,
\begin{equation}
	\delta \Phi^{\alpha(3)\dot{\alpha}(3)}
	=
	\partial^{(\alpha(\dot{\alpha}} \xi^{\alpha(2))\dot{\alpha}(2))},
	\qquad
	\delta \Phi^{\alpha\dot{\alpha}} =
	\partial_{\beta\dot{\beta }} \xi^{(\alpha\beta)(\dot{\alpha}\dot{\beta})},
\end{equation}
leads to the current conservation equation:
 \begin{equation}
 	\partial^{\beta\dot{\beta}} J_{(\alpha(2)\beta)(\dot{\alpha}(2)\dot{\beta})} + \partial_{(\alpha(\dot{\alpha}} T_{\alpha)\dot{\alpha})} \approx 0.
 \end{equation}

In the spinor formalism, Weyl-like tensors naturally decompose into (anti-)self-dual parts. 
Let $h_{\alpha(2)\dot{\alpha}(2)}$ and $A_{\alpha\dot{\alpha}} $
be the fields of spin-2 and spin-1 correspondingly. Then
 spin-2 and spin-1 self-dual tensors are given by:
\begin{equation}
	C_{\alpha(4)} = (\partial^2)^{\dot{\beta}(2)}_{(\alpha(2)} h_{\alpha(2))\dot{\beta}(2)},
	\qquad
	C_{\alpha(2)} = \partial_{(\alpha}^{\dot{\beta}}A_{\alpha)\dot{\beta}}.
\end{equation}	
 The anti-self-dual parts are obtained by the complex conjugation. On the equations of motion, Weyl-like tensors satisfy
 \begin{equation}
 	\partial_{\dot{\alpha}}^\beta C_{\beta\alpha(3)} \approx 0,
 	\qquad
 		\partial_{\dot{\alpha}}^\beta C_{\beta\alpha} \approx 0.
 \end{equation}	
 Using these spin-tensors, one can construct a complex spin-$3$ current with four derivatives as:
\begin{equation}\label{eq: spin 3 complex}
	\mathbf{J}_{\alpha(3)\dot{\alpha}(3)} = C_{\alpha(3)\beta} \partial^\beta_{(\dot{\alpha}} \bar{C}_{\dot{\alpha}(2))}.
\end{equation}	
In this order of derivatives, it is impossible to construct the current trace, meaning that $T_{\alpha\dot{\alpha}}=0$. Consequently, the trace partner $\Phi^{\alpha\dot{\alpha}}$
completely decouples from this interaction, and the vertex is governed solely by the first term in \eqref{eq: (321) vertex}.
This current \eqref{eq: spin 3 complex} satisfies the conservation equation:
\begin{equation}
	\partial^{\rho\dot{\rho}} \mathbf{J}_{(\rho\alpha(2))(\dot{\rho}\dot{\alpha}(2))}
	=
	\left(\partial^{\rho\dot{\rho}} C_{\rho\alpha(2)\beta}\right) \partial^\beta_{(\dot{\rho}} \bar{C}_{\dot{\alpha}(2))}
	+
	C_{\rho\alpha(2)\beta} \, \partial^{\rho\dot{\rho}} \left( \partial^\beta_{(\dot{\rho}} \bar{C}_{\dot{\alpha}(2))} \right)
	\approx 0.
\end{equation}

The current \eqref{eq: spin 3 complex} is complex, so it cannot be directly used to construct the cubic vertex \eqref{eq: (321) vertex}. However, one can consider the real and imaginary parts of this complex spin-3 current:
\begin{equation}
	\begin{split}
&\text{Re}\,	\mathbf{J}_{\alpha(3)\dot{\alpha}(3)} = \frac{1}{2}\left\{	C_{\alpha(3)\beta} \partial^\beta_{(\dot{\alpha}} \bar{C}_{\dot{\alpha}(2))} 
+
\partial_{(\alpha}^{\dot{\beta}} C_{\alpha(2))} \bar{C}_{\dot{\alpha}(3)\dot{\beta}}
\right\},
\\
&\text{Im}\,	\mathbf{J}_{\alpha(3)\dot{\alpha}(3)}  = \frac{1}{2i}\left\{	C_{\alpha(3)\beta} \partial^\beta_{(\dot{\alpha}} \bar{C}_{\dot{\alpha}(2))} 
-
\partial_{(\alpha}^{\dot{\beta}} C_{\alpha(2))} \bar{C}_{\dot{\alpha}(3)\dot{\beta}}
\right\}.
\end{split}
\end{equation}	
Each of these currents defines a consistent cubic vertex $(3,2,1)$:
\begin{equation}
		S^{\text{P-even}}_{(3,2,1)}
		=
		\int d^4x\, \Phi^{\alpha(3)\dot{\alpha}(3)} \text{Re}\,	\mathbf{J}_{\alpha(3)\dot{\alpha}(3)}, 
		\qquad
		S^{\text{P-odd}}_{(3,2,1)}
		=
			\int d^4x\, \Phi^{\alpha(3)\dot{\alpha}(3)} \text{Im}\,	\mathbf{J}_{\alpha(3)\dot{\alpha}(3)}.
\end{equation}	
Note that the vertex constructed from the imaginary part of the current is parity-breaking.

\medskip

We proceed to the $\mathcal{N}=2$ generalization of this vertex. According to the general construction, we seek the vertex  $(\mathbf{3}, \mathbf{2}, \mathbf{1})$ in the following form:
\begin{equation}\label{eq: SF (321) vertex}
	S_{(\mathbf{3}, \mathbf{2}, \mathbf{1})}
	=
	\int d^4x d^8\theta du \left( \Psi^{-(\alpha\beta)\dot{\alpha}} \mathcal{D}^+_\beta  
	+
	\bar{\Psi}^{-(\dot{\alpha}\dot{\beta})\alpha}
	\bar{\mathcal{D}}^+_{\dot{\beta}}
	\right)
	 \mathcal{J}_{\alpha\dot{\alpha}},
\end{equation}	
where $\mathcal{J}_{\alpha\dot{\alpha}}$ is a real principal supercurrent. Gauge transformations of the Mezincescu spin-$3$ prepotential are
\begin{equation}
		\begin{split}
			\delta \Psi^{-}_{(\alpha\beta)\dot{\alpha}}
			=
			\;
			&\mathcal{D}^{++} K^{(-3)}_{(\alpha\beta)\dot{\alpha} }
			+
			\mathcal{D}^+_{(\alpha} B^{--}_{\beta)\dot{\alpha}}
			+
			\mathcal{D}^{+\rho} B^{--}_{(\alpha\beta\rho)\dot{\alpha}}
			+
			\bar{\mathcal{D}}^{+\dot{\beta}} B^{--}_{(\alpha\beta)(\dot{\alpha}\dot{\beta})}
		+
			\bar{\mathcal{D}}^+_{\dot{\alpha}} B^{--}_{(\alpha\beta)} ,
		\end{split}
\end{equation}	
 and they imply that the principal supercurrent for vertex consistency must satisfy
\begin{equation}
	\begin{cases}
	\mathcal{D}^{++} \mathcal{J}_{\alpha\dot{\alpha}} \approx 0,
	\\
	\mathcal{D}^{+\alpha} \mathcal{J}_{\alpha\dot{\alpha}} \approx 0,
	\\
	\bar{\mathcal{D}}^{+\dot{\alpha}} \mathcal{J}_{\alpha\dot{\alpha}} \approx 0.
	\end{cases}
\end{equation}	 
Analogously to the previous example, to find its explicit structure, we first introduce the \textit{complex $\mathcal{N}=2$ principal spin-$3$ supercurrent}, using the chiral $\mathcal{N}=2$ super-Weyl tensor $\mathcal{W}_{(\alpha\beta)}$ and the chiral $\mathcal{N}=2$ Maxwell superfield strength $\mathcal{W}$ (see Appendix~\ref{app} for their definition):
\begin{equation}
\mathscr{J}_{\alpha\dot{\alpha}}
=
\mathcal{W}_{(\alpha\beta)} \partial^\beta_{\dot{\alpha}} \bar{\mathcal{W}}.
\end{equation}
This supercurrent satisfies the following equations:
\begin{equation}\label{eq: cons susy current}
	\begin{cases}
		\mathcal{D}^{++} \mathscr{J}_{\alpha\dot{\alpha}} = 0,
		\\
		\mathcal{D}^{+\alpha} \mathscr{J}_{\alpha\dot{\alpha}} \approx 0,
		\\
		\bar{\mathcal{D}}^{+\dot{\alpha}}\mathscr{J}_{\alpha\dot{\alpha}} \approx 0.
 	\end{cases}	
\end{equation}
To verify the conservation law \eqref{eq: cons susy current}, one can explicitly apply the spinor covariant derivatives  to the supercurrent and use properties of $\mathcal{N}=2$ super-Weyl tensors presented in Appendix \ref{app}. For example,
\begin{equation}\label{eq: explicit divergence}
	\mathcal{D}^{+\alpha} \mathscr{J}_{\alpha\dot{\alpha}} 
	= 
	\left( \mathcal{D}^{+\alpha} \mathcal{W}_{(\alpha\beta)} \right) \partial^\beta_{\dot{\alpha}} \bar{\mathcal{W}} 
	+ 
	\mathcal{W}_{(\alpha\beta)} \partial^\beta_{\dot{\alpha}} \left( \mathcal{D}^{+\alpha} \bar{\mathcal{W}} \right) 
	\approx 0.
\end{equation}
Similarly, for the conjugate spinorial derivative, the relation holds due to the off-shell chirality of the super-Weyl tensor ($\bar{\mathcal{D}}^{+\dot{\alpha}} \mathcal{W}_{(\alpha\beta)} = 0$) and the on-shell equations of motion for the Maxwell multiplet:
\begin{equation}\label{eq: explicit divergence conj}
	\bar{\mathcal{D}}^{+\dot{\alpha}} \mathscr{J}_{\alpha\dot{\alpha}} 
	= 
	\bar{\mathcal{D}}^{+\dot{\alpha}} \left( \mathcal{W}_{(\alpha\beta)} \partial^\beta_{\dot{\alpha}} \bar{\mathcal{W}} \right)
	=
	\mathcal{W}_{(\alpha\beta)} \partial^\beta_{\dot{\alpha}} \left( \bar{\mathcal{D}}^{+\dot{\alpha}} \bar{\mathcal{W}} \right) 
	\approx 0.
\end{equation}
The expression $\partial^\beta_{\dot{\alpha}} \bar{\mathcal{D}}^{+\dot{\alpha}} \bar{\mathcal{W}}$ vanishes on-shell, which physically corresponds to the Dirac equation for the photino component of the vector multiplet.

As in the bosonic example, one can also construct the interaction using the real or imaginary parts of the complex spin-$\mathbf{3}$ supercurrent. Accordingly, they will be responsible for the $\mathcal{N}=2$ supersymmetrization of the parity-invariant and the parity-breaking $(3,2,1)$ cubic interactions:
	\begin{equation}
		\begin{split}
	& \text{Re}	\mathscr{J}_{\alpha\dot{\alpha}}
		=
		\frac{1}{2}
		\left(
		\mathcal{W}_{(\alpha\beta)} \partial^\beta_{\dot{\alpha}} \bar{\mathcal{W}}
		+
		\partial_{\alpha}^{\dot{\beta}} \mathcal{W} \bar{\mathcal{W}}_{(\dot{\alpha}\dot{\beta})}
		\right),
\\&
	 \text{Im}	\mathscr{J}_{\alpha\dot{\alpha}}
		=
			\frac{1}{2i}
		\left(
		\mathcal{W}_{(\alpha\beta)} \partial^\beta_{\dot{\alpha}} \bar{\mathcal{W}}
		-
			\partial_{\alpha}^{\dot{\beta}} \mathcal{W} \bar{\mathcal{W}}_{(\dot{\alpha}\dot{\beta})}
		\right).
		\end{split}
	\end{equation}	
To see how the conservation equations \eqref{eq: cons susy current} guarantee the gauge invariance of the cubic vertex, one performs an integration by parts in the superfield action \eqref{eq: SF (321) vertex} under the gauge shifts of the prepotentials. For instance, the variation with respect to the harmonic parameter $K^{(-3)}_{(\alpha\beta)\dot{\alpha}}$ vanishes due to the harmonic constraint $\mathcal{D}^{++}\mathscr{J}_{\alpha\dot{\alpha}} = 0$. Similarly, the shifts proportional to the gauge parameters $B^{--}$ translate directly into the requirement for the supercurrent to be killed by the spinor covariant derivatives $\mathcal{D}^{+\alpha}$ and $\bar{\mathcal{D}}^{+\dot{\alpha}}$.
 
 Since both $\text{Re}\,\mathscr{J}_{\alpha\dot{\alpha}}$ and $\text{Im}\,\mathscr{J}_{\alpha\dot{\alpha}}$ dynamically satisfy the full set of these constraints on-shell, we conclude that \eqref{eq: SF (321) vertex} provides a consistent, gauge-invariant $\mathcal{N}=2$ superfield generalization of the four-derivative $(3,2,1)$ vertices.


\newpage

\section{General higher-spin currents }\label{sec: 3}

Let us now proceed to construct general component higher-spin currents. As in the example above, for the consistency of the vertex, the higher-spin current must satisfy a conservation condition:
\begin{equation}\label{eq: cons eq}
	\partial^{\beta\dot{\beta}} J_{(\beta \alpha(s-1))(\dot{\beta}\dot{\alpha}(s-1))}
	+
	\partial_{(\alpha(\dot{\alpha}} T_{\alpha(s-2))\dot{\alpha}(s-2))} \approx 0.
\end{equation}
The main building block for the currents is the Weyl-like tensor, defined by
\begin{equation}
	C_{\alpha(2s)} \coloneqq (\partial^s)^{\dot{\beta}(s)}_{(\alpha(s)} \Phi_{\alpha(s))\dot{\beta}(s)}.
\end{equation}
This is the only tensor that does not vanish on the equations of motion. On-shell, it satisfies
\begin{equation}
	\partial_{\dot{\alpha}}^\beta C_{\beta\alpha(2s-1)} \approx 0.
\end{equation}
We also use the condensed notation $C_{s}$ for the spin-$s$ Weyl-like tensor.

The general gauge-invariant ansatz for a complex spin-$s$ current  involving $s+s_1-s_2$ derivatives has the form\footnote{We use the following notation for the symmetrized product of derivatives: $\partial^p_{\alpha(p)\dot{\alpha}(p)}:= \partial_{(\alpha_1(\dot{\alpha}_1} \dots \partial_{\alpha_p)\dot{\alpha}_p)}$.}
\begin{equation}
	\mathbf{J}_{\alpha(s)\dot{\alpha}(s)}
	=
	\sum_{p=0}^{s-s_1-s_2} a_p  \left( \partial^p C_{s_1}\right)_{\alpha(p+s_1+s_2) \beta(s_1-s_2) \dot{\alpha}(p)}
	\left( \partial^{s-2s_2-p} \bar{C}_{s_2}  \right)^{\beta(s_1-s_2)}_{\alpha(s-s_1-s_2-p)\dot{\alpha}(s-p)}. 
\end{equation}	
Here $a_p$ are arbitrary complex coefficients. 
In the first and second terms, full symmetrization is carried out over all dotted and undotted indices.
The upper limit of the sum is 
$s-s_1-s_2$, because the second term must contain at least $s_1-s_2$ derivatives to match the index structure.

The non-zero trace of the current exists only when $s-s_1-s_2\geq 2$. The general ansatz is given~by 
\begin{equation}
	\begin{split}
	\mathbf{T}_{\alpha(s-2)\dot{\alpha}(s-2)}
	=
	\sum_{p=0}^{s-s_1-s_2-2}
	b_p 
	& \left( \partial^{p+1} C_{s_1}\right)_{\alpha(p+s_1+s_2) \beta(s_1-s_2+1) (\dot{\alpha}(p)\dot{\beta})}
	 \\
	 &\qquad\times
	\left( \partial^{s-2s_2-p-1} \bar{C}_{s_2}  \right)^{\beta(s_1-s_2+1)\dot{\beta}}_{\alpha(s-s_1-s_2-p-2)\dot{\alpha}(s-p-2)}. 
	\end{split}
\end{equation}	
In total, the ansatz for the current trace contains $s-s_1-s_2-1$ independent complex parameters~$b_p$.

The conservation equation \eqref{eq: cons eq} yields the recurrence relations for coefficients:
\begin{equation}\label{eq: bos equation}
	a_p \cdot \frac{s-s_1-s_2-p}{s} \cdot \frac{s-p}{s}
	+
	a_{p+1}\cdot  \frac{s_1+s_2+p+1}{s} \cdot \frac{p+1}{s}
	+
	b_p + b_{p-1} = 0.
\end{equation}	
In this equation, we assume the following boundary conditions:
\begin{equation}
\begin{split}
&a_p = 0 \quad \text{for}\quad p<0 \;\;\; \text{or}\;\;\;  p>s-s_1-s_2,
\\
&b_p = 0 \quad \text{for}\quad p<0 \;\;\;\text{or}\;\;\; p> s-s_1-s_2-2.
\end{split}
\end{equation}
The solutions of this equation depend on the choice of complex coefficients~$b_p$.

\medskip

\noindent $\bullet$ In the $b_p = 0$ case, we obtain a unique solution:
\begin{equation}\label{eq: (0)}
	a^{(0)}_p = (-1)^p \frac{\binom{s}{s_1+s_2+p} \binom{s}{p} }{\binom{s}{s_1+s_2}}.
\end{equation}	
This solution exists for any $s\geq s_1\geq s_2$ provided that $s\geq s_1+s_2$.
Corresponding traceless higher-spin current is given by the following expression:
\begin{equation}\label{eq: bos traceless}
	\begin{split}
	\mathbf{J}^{(0)}_{\alpha(s)\dot{\alpha}(s)}
	=
	\sum_{p=0}^{s-s_1-s_2} (-1)^p \frac{\binom{s}{s_1+s_2+p} \binom{s}{p} }{\binom{s}{s_1+s_2}}
&	  \left( \partial^p C_{s_1}\right)_{\alpha(p+s_1+s_2) \beta(s_1-s_2) \dot{\alpha}(p)}
\\
&\times	\left( \partial^{s-2s_2-p} \bar{C}_{s_2}  \right)^{\beta(s_1-s_2)}_{\alpha(s-s_1-s_2-p)\dot{\alpha}(s-p)}. 
	\end{split}
\end{equation}	

\noindent$\bullet$   Solutions with $b_p\neq 0$ are 
given by:
\begin{equation}\label{eq: gen sol}
a_p = a_p^{(0)} \left[ \frac{a_0}{a_0^{(0)}} + s^2 \sum_{j=0}^{p-1} \frac{b_j + b_{j-1}}{(s-s_1-s_2-j)(s-j) a_j^{(0)}} \right],
	\qquad
	p=1,2,\dots, s-s_1-s_2.
\end{equation}

In total, if $s-s_1-s_2\geq 2$, there are $s-s_1-s_2$ complex conserved currents; if $s-s_1-s_2 = 1$ or $s-s_1-s_2 = 0$, there is only one traceless complex current.

\medskip

We give two illustrative examples: one explains the general structure and relation to the ``fake'' vertices\footnote{The ``fake'' or trivial vertices are vertices  that can be eliminated by local field redefinitions.}, while the other one illustrates a reduction of number of independent real currents in the important limiting $s_1=s_2$ case.

\smallskip

\noindent \textbf{Example: $(5,2,1)$ vertex}

\smallskip

\noindent In this case, there are two complex currents. The traceless current is given by:
\begin{equation}
	\mathbf{J}^{(0)}_{\alpha(5)\dot{\alpha}(5)}
	=\;
	\bigl(C_2\bigr)_{\alpha(3)\beta}
	\bigl(\partial^3\bar C_1\bigr)^{\beta}_{\alpha(2)\dot{\alpha}(5)} 
	-\frac{5}{2}\;
	\bigl(\partial C_2\bigr)_{\alpha(4)\beta\dot{\alpha}}
	\bigl(\partial^2\bar C_1\bigr)^{\beta}_{\alpha\dot{\alpha}(4)} 
	+\;
	\bigl(\partial^2 C_2\bigr)_{\alpha(5)\beta\dot{\alpha}(2)}
	\bigl(\partial\bar C_1\bigr)^{\beta}_{\dot{\alpha}(3)}.
\end{equation}
The current with a nontrivial trace has the form:
\begin{subequations}
	\begin{equation}
		\mathbf{J}^{(1)}_{\alpha(5)\dot{\alpha}(5)} 
		=
		-   \frac{5^2}{4} \left( \partial C_2 \right)_{\alpha(4)\beta\dot{\alpha}}
		\left( \partial^2 \bar{C}_1 \right)^\beta_{\alpha\dot{\alpha}(4)},
	\end{equation}	
	\begin{equation}
		\mathbf{T}^{(1)}_{\alpha(3)\dot{\alpha}(3)}
		=
		 \left( \partial C_2 \right)_{\alpha(3)\beta(2)\dot{\beta}}
		\left( \partial^2 \bar{C}_1  \right)^{\beta(2)\dot{\beta}}_{\dot{\alpha}(3)}.
	\end{equation}	
\end{subequations}


After taking the real and imaginary parts of these currents, we obtain four real currents (two from each complex one). Not all of them give independent nontrivial cubic vertices, since some of vertices reduce to the ``fake'' interactions. In this example,  the "fake" $(5,2,1)$ vertices can be constructed as the real and imaginary parts of a complex gauge-invariant Lagrangian:
\begin{equation}\label{eq: fake (5,2,1)}
	\mathcal{L}^{\text{fake}}_{(5,2,1)}
	=
	\mathcal{R}^{\alpha(3)\dot{\alpha}(3)}_{(s=5)}
	C_{\alpha(3)\beta} \partial^\beta_{\dot{\alpha}} \bar{C}_{\dot{\alpha}(2)}.
\end{equation}	
Here $	\mathcal{R}^{\alpha(3)\dot{\alpha}(3)}_{(s=5)}$ is the spin-$5$ generalization of the linearized Ricci scalar, defined by \cite{Zaigraev:2026jvu}:
\begin{equation}\label{eq: curv 2}
	\begin{split}
		\mathcal{R}_{\alpha(3)\dot{\alpha}(3)}
		:=\;&
		\Box \Phi_{\alpha(3)\dot{\alpha}(3)}
		+
	2 \partial_{\alpha\dot{\alpha}} \partial^{\beta\dot{\beta}}\Phi_{(\alpha(2)\beta)(\dot{\alpha}(2)\dot{\beta})}
		-
		\frac{50}{9}  
		\partial^{\beta\dot{\beta}}\partial^{\gamma\dot{\gamma}} \Phi_{(\alpha(3)\beta\gamma)(\dot{\alpha}(3)\dot{\beta}\dot{\gamma})}.
	\end{split}
\end{equation}
Since the linearized Ricci tensor vanishes on-shell, $	\mathcal{R}_{\alpha(3)\dot{\alpha}(3)}\approx 0$, any vertex proportional to it can be systematically absorbed into the free action via local field redefinitions of the spin-$5$ field.  A ``fake'' vertex \eqref{eq: fake (5,2,1)} can also be written as: 
\begin{equation}\label{eq: fake (5,2,1)-2}
		\mathcal{L}^{\text{fake}}_{(5,2,1)}
	=
	-\frac{50}{4} \Phi^{\alpha(5)\dot{\alpha}(5)} \mathbf{J}^{(0)}_{\alpha(5)\dot{\alpha}(5)}
	+
	4 \left( \Phi^{\alpha(5)\dot{\alpha}(5)} \mathbf{J}^{(1)}_{\alpha(5)\dot{\alpha}(5)} + \Phi^{\alpha(3)\dot{\alpha}(3)}  	\mathbf{T}^{(1)}_{\alpha(3)\dot{\alpha}(3)} \right).
\end{equation}	

As a result, after taking into account the field redefinitions that lead to the ``fake'' vertices, as in the example of the $(3,2,1)$ vertex discussed in Section~\ref{sec: 2}, two nontrivial vertices, parity-invariant and parity-breaking, remain.  For the explicit construction of these vertices, one can take any combination of higher-spin currents that does not reduce to a fake vertex \eqref{eq: fake (5,2,1)-2}. For example, one can take the traceless higher-spin currents:
\begin{equation}
	\begin{split}
		S^{\text{P-even}}_{(5,2,1)}
		=
			\int d^4x\, \Phi^{\alpha(5)\dot{\alpha}(5)} \text{Re}\,	\mathbf{J}^{(0)}_{\alpha(5)\dot{\alpha}(5)},
		\qquad
		S^{\text{P-odd}}_{(5,2,1)}
		=
				\int d^4x\, \Phi^{\alpha(5)\dot{\alpha}(5)} \text{Im}\,	\mathbf{J}^{(0)}_{\alpha(5)\dot{\alpha}(5)}.
	\end{split}	
\end{equation}	

An absolutely analogous mechanism works in the general case as well.  The number of ``fake'' vertices coincides with the number of independent $a_p$ coefficients in the ansatz for the spin-$(s-2)$ current linear in Weyl-type tensors for spins $s_1$ and $s_2$ and is equal to $s-s_1-s_2-1$.
Such vertices are also constructed using the spin-$s$ linearized scalar curvature and have the form:
\begin{equation}
\mathcal{R}_{(s)}^{(s-2)\dot{\alpha}(s-2)}\left( \partial^p C_{s_1}\right)_{\alpha(p+s_1+s_2) \beta(s_1-s_2) \dot{\alpha}(p)}
\left( \partial^{s-2-2s_2-p} \bar{C}_{s_2}  \right)^{\beta(s_1-s_2)}_{\alpha(s-2-s_1-s_2-p)\dot{\alpha}(s-2-p)},
\end{equation}
where $p = 0,1, \dots,  s-s_1-s_2-2$.
With this taken into account, one nontrivial complex current remains, along with two nontrivial vertices corresponding to the parity-invariant and the parity-breaking $(s,s_1,s_2)$ interactions.

\medskip 

\noindent \textbf{Example: $(5,1,1)$ vertex}

\smallskip

\noindent The case $s_1=s_2$ is special, so we find it instructive to analyze. 
The traceless spin-$5$ current is given~by:
\begin{equation}\label{eq: spin 5-1}
	\begin{split}
		\mathbf{J}^{(0)}_{\alpha(5)\dot{\alpha}(5)} =
		&\;
	  (C_1)_{\alpha(2)} (\partial^3 \bar{C}_1)_{\alpha(3)\dot{\alpha}(5)}
		- 5 (\partial C_1)_{\alpha(3)\dot{\alpha}} (\partial^2 \bar{C}_1)_{\alpha(2)\dot{\alpha}(4)}
	\\&	\quad+
		5  (\partial^2 C_1)_{\alpha(4)\dot{\alpha}(2)} (\partial \bar{C}_1)_{\alpha\dot{\alpha}(3)} 
		-
		(\partial^3 C_1)_{\alpha(5)\dot{\alpha}(3)}  (\bar{C}_1)_{\dot{\alpha}(2)}.
	\end{split}
\end{equation}	
From the form of this current, it is clear that it is purely imaginary. This immediately indicates that there will be fewer conserved real currents in this case.

According to \eqref{eq: gen sol}, in this case there are two complex spin‑5 currents with the nonvanishing trace.  These currents are determined by  complex coefficients $b_0$ and $b_1$. It is convenient to represent the obtained current coefficients in the vector form:
\begin{equation}
\begin{pmatrix}
	 a_0 \\ a_1 \\ a_2 \\ a_3
	  \end{pmatrix}
= \alpha \begin{pmatrix} 1 \\ -5 \\ 5 \\ -1 \end{pmatrix}
+ b_0 \begin{pmatrix} 0 \\ -\frac{25}{3} \\ \frac{125}{24} \\ -\frac{25}{24} \end{pmatrix}
+ b_1 \begin{pmatrix} 0 \\ 0 \\ -\frac{25}{8} \\ -\frac{25}{24} \end{pmatrix},
\qquad \alpha, b_0, b_1 \in \mathbb{C}.
\end{equation}

Since the physical interaction Lagrangian is real, the complex currents must combine into the real structures. Therefore, we project the general solution onto the subspace of real currents and traces. The cross-symmetry under the exchange of $s_1\leftrightarrow s_2$	
dictates that the reality condition forces the trace parameters to be complex conjugates of each other:
\begin{equation}
	b_1^* = b_0 \quad \Leftrightarrow \quad b_0 = r + im, \quad b_1 = r - im, \quad (r, m \in \mathbb{R}).
\end{equation}
Extracting the independent components corresponding to the parameters $\text{Im}\, \alpha$, $m$, and $r$, we obtain exactly three linearly independent real spin-5 currents. To distinguish them from the complex precursors, we denote the real currents and their traces using regular font. In the expressions for $J^{(n)}$, the derivative indices are omitted for brevity and full symmetrization is implied:
\begin{equation}\label{eq: spin 5-0}
	\begin{split}
		J^{(0)}_{\alpha(5)\dot{\alpha}(5)} =
		&\;
		i
		\left(  C_{\alpha(2)} \partial^3 \bar{C}_{\dot{\alpha}(2)}
		- 5 \partial C_{\alpha(2)} \partial^2 \bar{C}_{\dot{\alpha}(2)}
		+
		5  \partial^2 C_{\alpha(2)} \partial \bar{C}_{\dot{\alpha}(2)} 
		-
		\partial^3 C_{\alpha(2)}  \bar{C}_{\dot{\alpha}(2)}
		\right),
	\end{split}
\end{equation}	
\begin{subequations}\label{eq: spin 5-2}
	\begin{equation}
		J^{(1)}_{\alpha(5)\dot{\alpha}(5)} =
		i \, \frac{25}{3}\left( \partial C_{\alpha(2)} \partial^2 \bar{C}_{\dot{\alpha}(2)} - \partial^2 C_{\alpha(2)} \partial \bar{C}_{\dot{\alpha}(2)} \right),
	\end{equation}	
	\begin{equation}
		T^{(1)}_{\alpha(3)\dot{\alpha}(3)}
		=	
		-i  \left( (\partial C)_{(\alpha(2)\beta)\dot{\beta}} (\partial^{2}\bar{C})^{\beta\dot{\beta}}_{\alpha\dot{\alpha}(3)} - (\partial^2 C)_{(\alpha(3)\beta)(\dot{\alpha}\dot{\beta})} (\partial \bar{C})^{\beta\dot{\beta}}_{\dot{\alpha}(2)}   \right).
	\end{equation}	
\end{subequations}
\begin{subequations}\label{eq: spin 5-3}
	\begin{equation}
		\begin{split}
			J^{(2)}_{\alpha(5)\dot{\alpha}(5)} =
			&\;
			-\frac{25}{24}
			\left(  C_{\alpha(2)} \partial^3 \bar{C}_{\dot{\alpha}(2)}
			+3 \partial C_{\alpha(2)} \partial^2 \bar{C}_{\dot{\alpha}(2)}
			+
			3  \partial^2 C_{\alpha(2)} \partial \bar{C}_{\dot{\alpha}(2)} 
			+
			\partial^3 C_{\alpha(2)}  \bar{C}_{\dot{\alpha}(2)}
			\right),
		\end{split}
	\end{equation}	
	\begin{equation}
		T^{(2)}_{\alpha(3)\dot{\alpha}(3)} 
		=
		(\partial C)_{(\alpha(2)\beta)\dot{\beta}} (\partial^{2}\bar{C})^{\beta\dot{\beta}}_{\alpha\dot{\alpha}(3)} + (\partial^2 C)_{(\alpha(3)\beta)(\dot{\alpha}\dot{\beta})} (\partial \bar{C})^{\beta\dot{\beta}}_{\dot{\alpha}(2)}.  
	\end{equation}	
\end{subequations}
This example illustrates how, in the case 
$s_1=s_2$, the number of independent real currents reduces compared to the generic case. Instead of the $2(s-s_1-s_2)$  independent real currents expected in the generic case, the symmetry dictates that only $s-2s_1$ currents survive.

The real ``fake'' cubic $(5,1,1)$ vertices  take the form:
\begin{subequations}
\begin{equation}
	\begin{split}
		\mathcal{L}^{\text{fake-I}}_{(5,1,1)}
		=\,&
		\mathcal{R}^{\alpha(3)\dot{\alpha}(3)}_{(s=5)} \Big\{C_{\alpha(2)} (\partial \bar{C})_{\alpha\dot{\alpha}(3)}
		+
		(\partial C)_{\alpha(3)\dot{\alpha}}  \bar{C}_{\dot{\alpha}(2)}
		 \Big\} 
		 \\
		 =\,& \frac{16}{3}   \Big\{ 
		 	\Phi^{\alpha(5)\dot{\alpha}(5)}  J^{(2)}_{\alpha(5)\dot{\alpha}(5)} 
		 +
		\Phi^{\alpha(3)\dot{\alpha}(3)}
		T^{(2)}_{\alpha(3)\dot{\alpha}(3)} 
		 \Big\}, 
		 \end{split}
\end{equation}
\begin{equation}
	\begin{split}
		\mathcal{L}^{\text{fake-II}}_{(5,1,1)}
		=\,&
		i	\mathcal{R}^{\alpha(3)\dot{\alpha}(3)}_{(s=5)} \Big\{ C_{\alpha(2)} (\partial \bar{C})_{\alpha\dot{\alpha}(3)}
		-
		(\partial C)_{\alpha(3)\dot{\alpha}}  \bar{C}_{\dot{\alpha}(2)}
		\Big\} 
		\\
			=\,&
			-
			\frac{50}{9} \Phi^{\alpha(5)\dot{\alpha}(5)} J^{(0)}_{\alpha(5)\dot{\alpha}(5)}
			-
			4 \Big\{  \Phi^{\alpha(5)\dot{\alpha}(5)} J^{(1)}_{\alpha(5)\dot{\alpha}(5)} 
			+
			\Phi^{\alpha(3)\dot{\alpha}(3)} T^{(1)}_{\alpha(3)\dot{\alpha}(3)} 
			 \Big\}.
	\end{split}
\end{equation}		
\end{subequations}
These equalities hold up to total derivatives and modulo terms that vanish on the spin-1 mass shell.

As a result, we conclude that there is only one nontrivial cubic $(5,1,1)$ vertex, which is P-odd.  We can take it in the form:
\begin{equation}
 S^{\text{P-odd}}_{(5,1,1)} = \int d^4x\, \Phi^{\alpha(5)\dot{\alpha}(5)}  
 J^{(0)}_{\alpha(5)\dot{\alpha}(5)}.
\end{equation}	
To construct the even vertex, we need at least two different spin-1 fields; see, for example, the discussion in  \cite{Zinoviev:2010cr}.

\subsection*{Conformal higher-spin currents}

The currents obtained in \eqref{eq: bos traceless} are divergence-free and traceless; therefore, they are conformal\footnote{We are grateful to the referee for drawing our attention to this point and for providing the reference to the work~\cite{Gelfond:2006be}.}.
Using these currents, one can construct cubic interaction vertices between the Fradkin–Tseytlin conformal higher-spin field of spin $s$ \cite{Fradkin:1985am} and massless higher-spin fields. Spin $s$ conformal field is described by the gauge field $ \Phi^{\alpha(s) \dot{\alpha}(s)}$
which obeys the gauge transformation law
\begin{equation}
	\delta \Phi^{\alpha(s) \dot{\alpha}(s)}
	=
	\partial^{(\alpha(\dot{\alpha}}\xi^{\alpha(s-1))\dot{\alpha}(s-1))},
\end{equation}
and enters the action as
\begin{equation}
	S^{\text{conf}}_{(s,s_1,s_2)} = \int d^4x\,
	\Phi^{\alpha(s) \dot{\alpha}(s)} \mathbf{J}^{(0)}_{\alpha(s)\dot{\alpha}(s)}.
\end{equation}
In the work \cite{Gelfond:2006be} (see eq. 4.6), class of conformal higher-spin currents were explicitly presented, which coincide with ours in the case $s_1=s_2$, $s\geq 2s_1$. These currents are a particular limiting case of the general current obtained in \cite{Gelfond:2006be}  (eq. (4.2)), which in our notation takes the form:
\begin{equation}
	J^{s_1,s_2,p}_{\alpha(2s_1+p)\dot{\alpha}(2s_2+p)}
	= 
	\sum_{j=0}^p \frac{\binom{2s_1+p}{2s_1+j} \binom{2s_2+p}{j}}{\binom{s_1+s_2+p}{s_1+s_2}}
	(\partial^j C_{s_1})_{\alpha(2s_1+j)\dot{\alpha}(j)}
	( \partial^{p-j} \bar{C}_{s_2})_{\alpha(p-j)\dot{\alpha}(2s_2+p-j)}. 
\end{equation}


\section{General $\mathcal{N}=2$ higher-spin supercurrents}\label{sec: 4}

To construct $\mathcal{N}=2$ supersymmetric vertices $(\mathbf{s}, \mathbf{s_1}, \mathbf{s_2})$,  we follow the approach of \cite{Zaigraev:2024ryg}. According to this approach, such vertices have a universal form:
\begin{equation}\label{eq: general vertex}
	S_{(\mathbf{s}, \mathbf{s_1}, \mathbf{s_2})}
	=
	\int d^4x d^8\theta du \left(\Psi^{-(\alpha(s-2)\beta) \dot{\alpha}(s-2)} \mathcal{D}^+_{\beta}
	+
	\bar{\Psi}^{-\alpha(s-2)(\dot{\alpha}(s-2)\dot{\beta})} \bar{\mathcal{D}}^+_{\dot{\beta}} 
	 \right)
	 \mathcal{J}_{\alpha(s-2)\dot{\alpha}(s-2)},
\end{equation}	
where $\Psi^{-\alpha(s-1)\dot{\alpha}(s-2)}$ is the spin-$\mathbf{s}$ Mezincescu-type higher-spin prepotential \cite{Buchbinder:2022vra, Zaigraev:2026jvu}, defined up to the gauge freedom:
\begin{equation}
	\begin{split}
		\delta_{\lambda,b} \Psi^{-}_{\alpha(s-1)\dot{\alpha}(s-2) }
		=
		\;
		&\mathcal{D}^{++} K^{(-3)}_{\alpha(s-1)\dot{\alpha}(s-2) }
		+
		\mathcal{D}^+_{(\alpha} B^{--}_{\alpha(s-2))\dot{\alpha}(s-2)}
		\\&+
		\mathcal{D}^{+\beta} B^{--}_{(\alpha(s-1)\beta)\dot{\alpha}(s-2)}
		+
		\bar{\mathcal{D}}^{+\dot{\beta}} B^{--}_{\alpha(s-1)(\dot{\alpha}(s-2)\dot{\beta})}
		\\
		&+
		\bar{\mathcal{D}}^+_{(\dot{\alpha}} B^{--}_{\alpha(s-1)\dot{\alpha}(s-3))}.
	\end{split}
\end{equation}
The superfield $\mathcal{J}_{\alpha(s-2)\dot{\alpha}(s-2)}$ is the real \textit{principal $\mathcal{N}=2$ supercurrent}, defined by the following equations:
\begin{equation}\label{eq: real principal supercurrent}
	\begin{cases}
		\widetilde{\mathcal{J}}_{\alpha(s-2)\dot{\alpha}(s-2)} = \mathcal{J}_{\alpha(s-2)\dot{\alpha}(s-2)},
		\\
		\mathcal{D}^{++} \mathcal{J}_{\alpha(s-2)\dot{\alpha}(s-2)} \approx 0,
		\\
		\mathcal{D}^{+\beta}\mathcal{J}_{(\beta\alpha(s-3))\dot{\alpha}(s-2)} \approx 0,
		\\
		\bar{\mathcal{D}}^{+\dot{\beta}} \mathcal{J}_{\alpha(s-2)(\dot{\beta}\dot{\alpha}(s-3))} \approx 0.
	\end{cases}	
\end{equation}	
If these conditions are satisfied, then the cubic interaction \eqref{eq: general vertex} is consistent at the leading order.

Following the construction of the spin-$\mathbf{3}$ supercurrent, we introduce a \textit{complex principal $\mathcal{N}=2$ supercurrent} $\mathscr{J}$ that satisfies the same equations \eqref{eq: real principal supercurrent}, except for the reality condition:
\begin{equation}\label{eq: complex principal supercurrent}
	\begin{cases}
		\mathcal{D}^{++} \mathscr{J}_{\alpha(s-2)\dot{\alpha}(s-2)} \approx 0,
		\\
		\mathcal{D}^{+\beta}\mathscr{J}_{(\beta\alpha(s-3))\dot{\alpha}(s-2)} \approx 0,
		\\
		\bar{\mathcal{D}}^{+\dot{\beta}} \mathscr{J}_{\alpha(s-2)(\dot{\beta}\dot{\alpha}(s-3))} \approx 0.
	\end{cases}	
\end{equation}	
Then, extracting the real and imaginary parts from the complex supercurrent
\begin{equation}
\begin{split}
	& \text{Re}\, \mathscr{J}_{\alpha(s-2)\dot{\alpha}(s-2)}  = \frac{1}{2} \left( \mathscr{J}_{\alpha(s-2)\dot{\alpha}(s-2)}  + \tilde{\mathscr{J}}_{\alpha(s-2)\dot{\alpha}(s-2)}  \right),
	\\
	& \text{Im}\, \mathscr{J}_{\alpha(s-2)\dot{\alpha}(s-2)}  = \frac{1}{2i} \left( \mathscr{J}_{\alpha(s-2)\dot{\alpha}(s-2)}  - \tilde{\mathscr{J}}_{\alpha(s-2)\dot{\alpha}(s-2)}  \right),
\end{split}	
\end{equation}
 we obtain the parity-even and the parity-odd real $\mathcal{N}=2$ supercurrents and the corresponding cubic vertices~\eqref{eq: general vertex}.

Before specifying the explicit form of the supercurrent, let us note that system \eqref{eq: complex principal supercurrent} inherently implies several general consequences; in particular:
\begin{subequations}
\begin{equation}
	\mathcal{D}^{--} \mathscr{J}_{(\beta\alpha(s-3))\dot{\alpha}(s-2)} \approx 0,
\end{equation}	
\begin{equation}
		\mathcal{D}^{-\beta}\mathscr{J}_{(\beta\alpha(s-3))\dot{\alpha}(s-2)} \approx 0,
\end{equation}	
\begin{equation}
		\bar{\mathcal{D}}^{-\dot{\beta}} \mathscr{J}_{\alpha(s-2)(\dot{\beta}\dot{\alpha}(s-3))} \approx 0,
\end{equation}	
\begin{equation}\label{eq: cons eq 0}
	\partial^{\beta\dot{\beta}} \mathscr{J}_{(\beta\alpha(s-3))(\dot{\beta}\dot{\alpha}(s-3))} \approx 0.
\end{equation}	
\end{subequations}
The last equation ensures that all component fields are traceless conserved currents (see Section~\ref{sec: 2} and eq.~\eqref{eq: bos traceless} above).

We proceed to construct a general spin-$\mathbf{s}$ complex $\mathcal{N}=2$ principal supercurrent linear in the spin-$\mathbf{s_1}$ and spin-$\mathbf{s_2}$ super-Weyl  tensors.  Such a construction is possible provided that $\mathbf{s} \geq \mathbf{s_1} + \mathbf{s_2}$. 
We start with an ansatz for the $\mathcal{N}=2$ complex principal supercurrent. For definiteness, we will  consider~$\mathbf{s_1} \geq \mathbf{s_2}$.

\smallskip

To systematically solve the constraints, we structure the ansatz for the complex principal supercurrent by grouping terms according to the number of supersymmetric covariant derivatives. Specifically, the coefficients $a_p$, $b_p$, and $c_p$ parameterize contributions containing zero, one, and two spinor derivatives acting on the super-Weyl tensor, respectively.
The most general ansatz then has the form:
\begin{equation}
	\begin{split}
&	\mathscr{J}_{\alpha(s-2)\dot{\alpha}(s-2)} 
	=
	\sum_{p=0}^{s-s_1-s_2}
	a_p
	\left(\partial^p
	\mathcal{W}\right)_{\alpha(s_1+s_2+p-2)\beta(s_1-s_2)\dot{\alpha}(p)}
	\left( \partial^{s-2s_2-p}\right)^{\beta(s_1-s_2)}_{\alpha(s-s_1-s_2-p)\dot{\alpha}(s-2s_2- p )} \bar{\mathcal{W}}_{\dot{\alpha}(2s_2-2)}
	\\
	&
	+	\sum_{p=0}^{s-s_1-s_2-1}
	b_p
	\Bigg[ \left(	\partial^p
	\mathcal{D}^+
	\mathcal{W}\right)_{\alpha(s_1+s_2+p-1)\beta(s_1-s_2)\dot{\alpha}(p)}
	\left( \partial^{s-2s_2-p-1}\right)^{\beta(s_1-s_2)}_{\alpha(s-s_1-s_2-p-1)\dot{\alpha}(s-2s_2- p -1 )}
	\bar{\mathcal{D}}^-_{\dot{\alpha}} \bar{\mathcal{W}}_{\dot{\alpha}(2s_2-2)}
	\\&
	\qquad\qquad\qquad
	-
	\left(\partial^p
	\mathcal{D}^-
	\mathcal{W}\right)_{\alpha(s_1+s_2+p-1)\beta(s_1-s_2)\dot{\alpha}(p)}
	\left( \partial^{s-2s_2-p-1}\right)^{\beta(s_1-s_2)}_{\alpha(s-s_1-s_2-p-1)\dot{\alpha}(s-2s_2- p-1 )}
	\bar{\mathcal{D}}^+_{\dot{\alpha}} \bar{\mathcal{W}}_{\dot{\alpha}(2s_2-2)}
	\Bigg]
		\\
	&
	+	\sum_{p=0}^{s-s_1-s_2-2}
	c_p
	\left(\partial^p
	\mathcal{D}^+	\mathcal{D}^-
	\mathcal{W}\right)_{\alpha(s_1+s_2+p)\beta(s_1-s_2)\dot{\alpha}(p)}
	\left( \partial^{s-2s_2-p-2}\right)^{\beta(s_1-s_2)}_{\alpha(s-s_1-s_2-p-2)\dot{\alpha}(s-2s_2- p-2)}
	\bar{\mathcal{D}}^+_{\dot{\alpha}}
	\bar{\mathcal{D}}^-_{\dot{\alpha}} \bar{\mathcal{W}}_{\dot{\alpha}(2s_2-2)}
	.
	\end{split}
\end{equation}	
Here, we use spin-$\mathbf{s_1}$ and spin-$\mathbf{s_2}$ Weyl-like supertensors $\mathcal{W}_{\alpha(2s_1-2)}$ and $\bar{\mathcal{W}}_{\dot{\alpha}(2s_2-2)}$.
In each term, we assume full symmetrization over all dotted and undotted indices.

This ansatz automatically satisfies the harmonic independence condition off-shell:
\begin{equation}
	\mathcal{D}^{++} \mathscr{J}_{\alpha(s-2)\dot{\alpha}(s-2)}  = 0.
\end{equation}	
Imposing the remaining conservation equations \eqref{eq: complex principal supercurrent} on the introduced ansatz leads to the set of coupled algebraic recurrence relations for coefficients. In particular, the undotted spinor derivative constraint yields
	\begin{equation}\label{eq: sys 1}
	\mathcal{D}^{+\beta}\mathscr{J}_{(\beta\alpha(s-3))\dot{\alpha}(s-2)} \approx 0
	\quad
	\Rightarrow
	\quad
	\begin{cases}
		a_p  (s-s_1-s_2-p)
		-
		4i\,
		b_p (s_1+s_2+p-	1)
		=
		0,
		\\
		b_p  (s-s_1-s_2-p-1)
		+
		4i
		\, c_p  (s_1+s_2+p)= 0.
	\end{cases}	
\end{equation}	
The action of the conjugate derivative gives additional constraints:
\begin{equation}\label{eq: sys 2}
		\bar{\mathcal{D}}^{+\dot{\beta}} \mathscr{J}_{\alpha(s-2)(\dot{\beta}\dot{\alpha}(s-3))} \approx 0
		\quad
		\Rightarrow
		\quad
		\begin{cases}
			a_p\, p
			+
			4i \, b_{p-1} (s-p-1) = 0,
			\\
			b_p\,  p
			-
			4i  \, c_{p-1}  (s-p-2)= 0.
		\end{cases}	
\end{equation}	
In addition, the general ansatz requires boundary conditions to be imposed on the coefficients:
\begin{equation}
	\begin{split}
		&a_p = 0 \quad \text{for}\quad p<0 \;\;\; \text{or}\;\;\;  p>s-s_1-s_2,
		\\
		&b_p = 0 \quad \text{for}\quad p<0 \;\;\;\text{or}\;\;\; p> s-s_1-s_2-1,
		\\
		&c_p = 0 \quad \text{for}\quad p<0 \;\;\;\text{or}\;\;\; p> s-s_1-s_2-2.
	\end{split}
\end{equation}

Solving the resulting equations \eqref{eq: sys 1} and \eqref{eq: sys 2}, we obtain the unique solution:
\begin{subequations}
\begin{equation}
	a_p =  (-1)^p   \frac{\binom{s-2}{p} \binom{s-2}{s_1+s_2+p-2} }{\binom{s-2}{s_1+s_2-2} },
	\qquad
	p = 0,1,2 \dots, s - s_1 - s_2,
\end{equation}	
\begin{equation}
	\qquad b_p = \frac{ (-1)^p}{4i}   \frac{\binom{s-2}{p} \binom{s-2}{s_1+s_2+p-1} }{\binom{s-2}{s_1+s_2-2} },
	\qquad
	p = 0,1,2 \dots, s - s_1 - s_2-1,
\end{equation}	
\begin{equation}
	\qquad \quad c_p = \frac{(-1)^{p}}{16}   \frac{\binom{s-2}{p} \binom{s-2}{s_1+s_2+p} }{\binom{s-2}{s_1+s_2-2} },
	\quad\quad
	p = 0,1,2 \dots, s - s_1 - s_2-2.
\end{equation}	
\end{subequations}


We conclude this section by listing some features of the constructed $\mathcal{N}=2$ complex principal supercurrents.

\begin{itemize}
\item  In the saturated spin limit $\mathbf{s}=\mathbf{s_1}+\mathbf{s_2}$, the structure of the supercurrent simplifies significantly, as the boundary conditions force the coefficients $b_p$ and $c_p$
to vanish identically, leaving only the single non-vanishing series parameterized by $a_0$:
\begin{equation}
\mathscr{J}_{\alpha(s-2)\dot{\alpha}(s-2)}=	\mathcal{W}_{\alpha(s_1+s_2-2)\beta(s_1-s_2)}
	\left( \partial^{s_1-s_2}\right)^{\beta(s_1-s_2)}_{\dot{\alpha}(s_1-s_2)} \bar{\mathcal{W}}_{\dot{\alpha}(2s_2-2)}.
\end{equation}	

\item In the equal-spin limit $\mathbf{s_1}= \mathbf{s_2}$, we reproduce the $\mathcal{N}=2$ principal supercurrents previously obtained in \cite{Zaigraev:2024ryg}. For odd $\mathbf{s}$, these supercurrents are purely imaginary, while for even $\mathbf{s}$, they are real.

\item The conditions \eqref{eq: complex principal supercurrent} defining the complex principal supercurrent lead to the conclusion that all component currents are conserved \eqref{eq: cons eq 0}. This does not mean that after component reduction all resulting vertices will be constructed from the traceless higher-spin currents \eqref{eq: bos traceless}. The reason for this lies in the fact that some of fields in the Wess–Zumino gauge correspond to traces of the higher-spin Lorentz connection and lead to the appearance of trace-type terms. For details, see~\cite{Zaigraev:2026jvu}.

\item In the bosonic case, there is a family of conserved higher-spin currents given by the equations~\eqref{eq: (0)} and \eqref{eq: gen sol}.  Our construction shows that, in the $\mathcal{N}=2$ case, there is only one principal higher-spin supercurrent. In components, the complex principal supercurrent contains a special combination of currents. For example, for the spin-$\mathbf{5}$ supercurrent constructed from spin $\mathbf{1}$, such a combination has the form:
\begin{equation}
	J_{\alpha(5)\dot{\alpha}(5)} = 4 J^{(0)}_{\alpha(5)\dot{\alpha}(5)} 
	+
	\frac{6}{25} J^{(1)}_{\alpha(5)\dot{\alpha}(5)}. 
\end{equation}	

\end{itemize}

\section{Conclusions} \label{sec: 5}

In this article, we have systematically investigated cubic interaction vertices for $\mathcal{N}=2$ higher-spin gauge supermultiplets in harmonic superspace and corresponding $\mathcal{N}=2$ supercurrents. As a simple example, we constructed the four-derivative $(\mathbf{3},\mathbf{2},\mathbf{1})$ superfield vertex, which provides the $\mathcal{N}=2$ extension of the component $(3,2,1)$ interaction.

Besides the supersymmetric analysis, we carried out the complete classification of the corresponding bosonic higher-spin currents, associated with the abelian $(s,s_1,s_2)$ interactions. Solving the current conservation equations, we obtained the complete family of complex currents, including currents with the non-vanishing trace, and identified the corresponding trivial (``fake'') vertices removable by local field redefinitions. After factoring out such vertices, a single nontrivial complex current remains in the generic case $s_1\neq s_2$, generating the parity-even and the parity-odd cubic interactions. We also analyzed the special case $s_1=s_2$, where additional reality conditions reduce the number of independent conserved currents.

For arbitrary $\mathcal{N}=2$ spins, we constructed the most general gauge-invariant ansatz for the complex principal supercurrent built from spin-$\mathbf{s_1}$ and spin-$\mathbf{s_2}$ Weyl-like supertensors with the minimal number of derivatives. Solving the supercurrent conservation equations, we found the unique exact solution, which exists provided the bound $\mathbf{s}\geq \mathbf{s_1}+\mathbf{s_2}$ is satisfied. In the saturated case $\mathbf{s}=\mathbf{s_1}+\mathbf{s_2}$, the supercurrent reduces to a particularly simple form.

Finally, we demonstrated that for $\mathbf{s_1}\neq\mathbf{s_2}$ the principal supercurrent is intrinsically complex and generates two independent cubic vertices whose component expansions contain the parity-even and the parity-odd $(s,s_1,s_2)$ interactions. In the equal-spin limit $\mathbf{s_1}=\mathbf{s_2}$, the supercurrent becomes purely imaginary for odd $\mathbf{s}$ and purely real for even $\mathbf{s}$, reproducing the results of \cite{Zaigraev:2024ryg, Zaigraev:2026jvu}.

\acknowledgments

The author is grateful to Yu.~Zinoviev for discussions of cubic vertices
and E.~Ivanov for many fruitful discussions and permanent support. 
This work was partially supported by
the Foundation for the Advancement of Theoretical Physics and
Mathematics ``BASIS'', grant \verb|#| 25-1-1-10-4 .

\appendix

\section{$\mathcal{N}=2$ Weyl-like higher-spin tensors }\label{app}

The only gauge-invariant object that does not vanish on the higher-spin equations of motion is the Weyl-like tensor. In this appendix, we  present the main details of its construction in the harmonic approach \cite{Ivanov:2024gjo, Ivanov:2025gvs, Ivanov:2025mld, Zaigraev:2026jvu}. For clarity, we will divide the construction into several stages.

\smallskip 

 \noindent \textbf{1.} \textbf{Supersymmetry-invariant prepotentials}
 
 \smallskip
 
 \noindent It is convenient to define superfields that are invariant under supersymmetry transformations. To this end, we consider the expansion of the spin-$\mathbf{s}$ differential operator \eqref{eq: diff operator} in the basis of covariant derivatives: 
 \begin{equation}\label{eq: DO cov G}
 	\hat{\mathcal{H}}^{++}_{(s)} = G^{++\alpha(s-2)\dot{\alpha}(s-2)M} \mathcal{D}_M \partial^{s-2}_{\alpha(s-2)\dot{\alpha}(s-2)},
 	\qquad
 	\delta_\epsilon G^{++\alpha(s-2)\dot{\alpha}(s-2)M} = 0.
 \end{equation}	
 Here, the covariant derivatives are given by
 $
 \mathcal{D}_M = \left\{ \partial_{\alpha\dot{\alpha}},
 \mathcal{D}^-_\alpha,
 \bar{\mathcal{D}}^-_{\dot{\alpha}}, \partial_5 \right\}
 $. The $\mathcal{D}^-$ covariant derivatives are defined by:
 \begin{equation*}
 	\begin{split}
 \mathcal{D}^-_\alpha &= - \partial^-_\alpha + 4i \bar{\theta}^{-\dot{\alpha}}\partial_{\alpha\dot{\alpha}}
 -
 2i \theta^-_\alpha \partial_5,
 \\
 \bar{\mathcal{D}}^-_{\dot{\alpha}} &= - \partial^-_{\dot{\alpha}} - 4i \theta^{-\alpha}\partial_{\alpha\dot{\alpha}}
 -
 2i \bar{\theta}^-_{\dot{\alpha}} \partial_5.
 \end{split}
 \end{equation*}
 
 \smallskip 
 
  \noindent \textbf{2.} \textbf{Composite  superfields $\mathcal{H}^{\pm\pm}_{\alpha(2s-2)}$}
 
 \smallskip
 
 \noindent Using supersymmetry-invariant superfields $G^{++}$, we can construct the composite superfield:
 \begin{equation}\label{eq: H++}
 	\mathcal{H}^{++}_{\alpha(2s-2)}
 	:=
 	\partial_{(\alpha_1}^{\;\;\dot{\beta}_1} \dots  	\partial_{\alpha_{s-2}}^{\;\;\dot{\beta}_{s-2}} \left(
 	\mathcal{D}^-_{\alpha_{s-1}} G^{+++}_{\alpha_s\dots\alpha_{2s-2})(\dot{\beta}_1\dots\dot{\beta}_{s-2})}
 	+
 	\partial_{\alpha_{s-1}}^{\;\;\dot{\beta}_{s-1}} G^{++}_{\alpha_s\dots\alpha_{2s-2})(\dot{\beta}_1\dots\dot{\beta}_{s-1})}\right).
 \end{equation}	
The key property possessed by this superfield is the ``half-analyticity'' condition \cite{Ivanov:2024gjo}:
 \begin{equation}\label{eq: H-A condition}
 	\bar{\mathcal{D}}^+_{\dot{\alpha}}\mathcal{H}^{++}_{\alpha(2s-2)} = 0.
 \end{equation}	
 It completely fixes the relative coefficient in \eqref{eq: H++}.
 Using the harmonic zero-curvature condition, one can construct a superfield with a negative harmonic charge:
  \begin{equation}\label{eq: ZC hcal}
 	\mathcal{D}^{++} \mathcal{H}^{--}_{\alpha(2s-2)} =  	\mathcal{D}^{--} \mathcal{H}^{++}_{\alpha(2s-2)}. 
 \end{equation}	
Thus, the obtained superfields have a simple gauge transformation law with the composite ``half-analytic'' parameter:
 \begin{equation}
 	\delta_\lambda \mathcal{H}^{\pm\pm}_{\alpha(2s-2)} 
 	=
 	\mathcal{D}^{\pm\pm} \Lambda_{\alpha(2s-2)},
 	\qquad 
 	\bar{\mathcal{D}}^+_{\dot{\alpha}} \Lambda_{\alpha(2s-2)} = 0.
 \end{equation}	
 
 \smallskip 
 
 \noindent \textbf{3.} \textbf{Higher-spin Weyl-like supertensor}
 
 \smallskip
 
\noindent We define the gauge-invariant superfield:
 \begin{equation}\label{eq: app W}
	\mathcal{W}_{\alpha(2s-2)} :=\left(\bar{\mathcal{D}}^+\right)^2 \mathcal{H}^{--}_{\alpha(2s-2)},
	\qquad
	\delta_\lambda 	\mathcal{W}_{\alpha(2s-2)} = 0.
\end{equation}	
Gauge invariance is ensured by the ``half-analyticity'' condition of the gauge parameter. Moreover, ``half-analyticity'' implies that the constructed superfield satisfies the conditions of harmonic independence and chirality:
 \begin{equation}\label{eq: W harm indep}
	\mathcal{D}^{\pm\pm} \mathcal{W}_{\alpha(2s-2)} = 0,
	\qquad
	\bar{\mathcal{D}}^\pm_{\dot{\alpha}} \mathcal{W}_{\alpha(2s-2)}  = 0.
\end{equation}	
In components, this superfield contains the higher-spin Weyl-like tensors, so it is natural to identify it as the $\mathcal{N}=2$ spin-$\mathbf{s}$ Weyl-like supertensor. For more details on the component structure see~\cite{Zaigraev:2026jvu}.

On-shell, the supertensor satisfies many useful properties. We will present the main ones:
\begin{equation}\label{eq: SW prop}
	\mathcal{D}^{\pm\alpha}  
	\mathcal{W}_{\alpha(2s-2)} \approx 0,
	\qquad
	\partial_{\dot{\alpha}}^{\;\alpha} \mathcal{W}_{\alpha(2s-2)} \approx 0,
	\qquad 
	\Box \mathcal{W}_{\alpha(2s-2)} \approx 0.
\end{equation}

Since for $s=1$ the superfield $\mathcal{W}_{\alpha(2s-2)}$
completely loses its spinor indices, this case requires separate consideration.
The superstrength is defined as $\mathcal{W}:= (\bar{\mathcal{D}}^+)^2 V^{--}$ and on-shell satisfies
\begin{equation}
	(\mathcal{D}^+)^2 \mathcal{W} \approx 0,
	\qquad
	\mathcal{D}^{+\alpha}\partial_\alpha^{\dot{\beta}} \mathcal{W} \approx 0.
\end{equation}



\begin{thebibliography}{99}
	

\bibitem{Metsaev:2005ar}
R.~R.~Metsaev,
{\it Cubic interaction vertices of massive and massless higher spin fields},
\href{https://doi.org/10.1016/j.nuclphysb.2006.10.002}{Nucl. Phys. B \textbf{759} (2006), 147-201}
[\href{https://arxiv.org/abs/0512342}{arXiv:hep-th/0512342 [hep-th]}].

\bibitem{Metsaev:2007rn}
R.~R.~Metsaev,
{\it Cubic interaction vertices for fermionic and bosonic arbitrary spin fields},
\href{https://doi.org/10.1016/j.nuclphysb.2012.01.022}{Nucl. Phys. B \textbf{859} (2012), 13-69}
[\href{https://arxiv.org/abs/0712.3526}{arXiv:0712.3526 [hep-th]}].

\bibitem{Berends:1985xx}
F.~A.~Berends, G.~J.~H.~Burgers and H.~van Dam,
{\it Explicit Construction of Conserved Currents for Massless Fields of Arbitrary Spin},
\href{https://doi.org/10.1016/S0550-3213(86)80019-0}{Nucl. Phys. B \textbf{271} (1986), 429-441}.


\bibitem{Gelfond:2006be}
O.~A.~Gelfond, E.~D.~Skvortsov and M.~A.~Vasiliev,
{\it Higher spin conformal currents in Minkowski space},
\href{https://doi.org/10.1007/s11232-008-0027-6}{Theor. Math. Phys. \textbf{154} (2008), 294-302}
[\href{https://arxiv.org/abs/hep-th/0601106}{arXiv:hep-th/0601106 [hep-th]}].

\bibitem{Manvelyan:2009vy}
R.~Manvelyan, K.~Mkrtchyan and W.~Ruhl,
{\it Off-shell construction of some trilinear higher spin gauge field interactions},
\href{https://doi.org/10.1016/j.nuclphysb.2009.07.007}{Nucl. Phys. B \textbf{826} (2010), 1-17}
[\href{https://arxiv.org/abs/0903.0243}{arXiv:0903.0243 [hep-th]}].

\bibitem{Zinoviev:2010cr}
Y.~M.~Zinoviev,
{\it Spin 3 cubic vertices in a frame-like formalism},
\href{https://doi.org/10.1007/JHEP08(2010)084}{JHEP \textbf{08} (2010), 084}
[\href{https://arxiv.org/abs/1007.0158}{arXiv:1007.0158 [hep-th]}].






\bibitem{Vasiliev:1990en}
M.~A.~Vasiliev,
{\it Consistent equation for interacting gauge fields of all spins in (3+1)-dimensions},
\href{https://doi.org/10.1016/0370-2693(90)91400-6}{Phys. Lett. B \textbf{243} (1990), 378-382}

\bibitem{Vasiliev:1992av}
M.~A.~Vasiliev,
{\it More on equations of motion for interacting massless fields of all spins in $(3+1)$-dimensions},
\href{https://doi.org/10.1016/0370-2693(92)91457-K}{Phys. Lett. B \textbf{285} (1992), 225-234}.



\bibitem{Gelfond:2017wrh}
O.~A.~Gelfond and M.~A.~Vasiliev,
{\it Current Interactions from the One-Form Sector of Nonlinear Higher-Spin Equations},
\href{https://doi.org/10.1016/j.nuclphysb.2018.04.017}{Nucl. Phys. B \textbf{931} (2018), 383-417}
[\href{https://arxiv.org/abs/1706.03718}{arXiv:1706.03718 [hep-th]}].


\bibitem{Misuna:2017bjb}
N.~Misuna,
{\it On current contribution to Fronsdal equations},
\href{https://doi.org/10.1016/j.physletb.2018.01.019}{Phys. Lett. B \textbf{778} (2018), 71-78}
[\href{https://arxiv.org/abs/1706.04605}{arXiv:1706.04605 [hep-th]}].

\bibitem{Tatarenko:2024csa}
Y.~A.~Tatarenko and M.~A.~Vasiliev,
{\it Bilinear Fronsdal currents in the AdS$_{4}$ higher-spin theory},
\href{https://doi.org/10.1007/JHEP07(2024)246}{JHEP \textbf{07} (2024), 246}
[\href{https://arxiv.org/abs/2405.02452}{arXiv:2405.02452} [hep-th]].

\bibitem{Gates:1983nr}
S.~J.~Gates, M.~T.~Grisaru, M.~Rocek and W.~Siegel,
{\it Superspace Or One Thousand and One Lessons in Supersymmetry},
Front. Phys. \textbf{58} (1983), 1-548
ISBN 978-0-8053-3161-5
[\href{https://arxiv.org/abs/hep-th/0108200}{arXiv:hep-th/0108200 [hep-th]}].

\bibitem{BK}
I. L. Buchbinder, S. M. Kuzenko, 
{\it Ideas and Methods of Supersymmetry and Supergravity or a Walk Through Superspace}, IOP Publishing, Bristol U.K., 1998.	

\bibitem{Wess:1992cp}
J.~Wess and J.~Bagger,
{\it Supersymmetry and supergravity},
Princeton University Press, 1992.

\bibitem{18} A.~S.~Galperin, E.~A.~Ivanov, V.~I.~Ogievetsky, E.~S.~Sokatchev,
{\it Harmonic superspace}, Cambridge Monographs on Mathematical
Physics, Cambridge University Press, 2001, 306 p.


\bibitem{Kuzenko:2017ujh}
S.~M.~Kuzenko, R.~Manvelyan and S.~Theisen,
{\it Off-shell superconformal higher spin multiplets in four dimensions},
\href{https://doi.org/10.1007/JHEP07(2017)034}{JHEP \textbf{07} (2017), 034}
[\href{https://arxiv.org/abs/1701.00682}{arXiv:1701.00682 [hep-th]}].

\bibitem{Hutomo:2017phh}
J.~Hutomo and S.~M.~Kuzenko,
{\it Non-conformal higher spin supercurrents},
\href{https://doi.org/10.1016/j.physletb.2018.01.045}{Phys. Lett. B \textbf{778} (2018), 242-246}
[\href{https://arxiv.org/abs/1710.10837}{arXiv:1710.10837 [hep-th]}].

\bibitem{Koutrolikos:2017qkx}
K.~Koutrolikos, P.~Ko{\v{c}}{\'\i} and R.~von Unge,
{\it Higher Spin Superfield interactions with Complex linear Supermultiplet: Conserved Supercurrents and Cubic Vertices},
\href{https://doi.org/10.1007/JHEP03(2018)119}{JHEP \textbf{03} (2018), 119}
[\href{https://arxiv.org/abs/1712.05150}{arXiv:1712.05150 [hep-th]}].

\bibitem{Buchbinder:2018wwg}
I.~L.~Buchbinder, S.~J.~Gates and K.~Koutrolikos,
{\it Interaction of supersymmetric nonlinear sigma models with external higher spin superfields via higher spin supercurrents},
\href{https://doi.org/10.1007/JHEP05(2018)204}{JHEP \textbf{05} (2018), 204}
[\href{https://arxiv.org/abs/1804.08539}{arXiv:1804.08539 [hep-th]}].

\bibitem{Buchbinder:2018gle}
I.~L.~Buchbinder, S.~J.~Gates and K.~Koutrolikos,
{\it Integer superspin supercurrents of matter supermultiplets},
\href{https://doi.org/10.1007/JHEP05(2019)031}{JHEP \textbf{05} (2019), 031}
[\href{https://arxiv.org/abs/1811.12858}{arXiv:1811.12858 [hep-th]}].

\bibitem{Buchbinder:2022kzl}
I.~Buchbinder, E.~Ivanov and N.~Zaigraev,
{\it Off-shell cubic hypermultiplet couplings to $ \mathcal{N} $ = 2 higher spin gauge superfields},
\href{https://doi.org/10.1007/JHEP05(2022)104}{JHEP \textbf{05} (2022), 104}
[\href{https://arxiv.org/abs/2202.08196}{arXiv:2202.08196 [hep-th]}].

\bibitem{Buchbinder:2022vra}
I.~Buchbinder, E.~Ivanov and N.~Zaigraev,
{\it $ \mathcal{N} $ = 2 higher spins: superfield equations of motion, the hypermultiplet supercurrents, and the component structure},
\href{doi:10.1007/JHEP03(2023)036}{JHEP \textbf{03} (2023), 036}
[\href{https://arxiv.org/abs/2212.14114}{arXiv:2212.14114 [hep-th]}].

\bibitem{Kuzenko:2023vgf}
S.~M.~Kuzenko and E.~S.~N.~Raptakis,
{\it On higher-spin $ \mathcal{N} = 2$ supercurrent multiplets},
\href{https://doi.org/10.1007/JHEP05(2023)056}{JHEP \textbf{05} (2023), 056}
[\href{https://arxiv.org/abs/2301.09386}{arXiv:2301.09386 [hep-th]}].


\bibitem{Buchbinder:2024pjm}
I.~Buchbinder, E.~Ivanov and N.~Zaigraev,
{\it $\mathcal{N}=2$ superconformal higher-spin multiplets and their hypermultiplet couplings},
\href{https://doi.org/10.1007/JHEP08(2024)120}{JHEP \textbf{08} (2024), 120}
[\href{https://arxiv.org/abs/2404.19016}{arXiv:2404.19016 [hep-th]}].

\bibitem{Kuzenko:2024vms}
S.~M.~Kuzenko and E.~S.~N.~Raptakis,
{\it Towards $ \mathcal{N} = 2$ superconformal higher-spin theory},
\href{https://doi.org/10.1007/JHEP11(2024)013}{JHEP \textbf{11} (2024), 013}
[\href{https://arxiv.org/abs/2407.21573}{arXiv:2407.21573 [hep-th]}].




	
	\bibitem{Buchbinder:2018wzq}
	I.~L.~Buchbinder, S.~J.~Gates and K.~Koutrolikos,
	{\it Conserved higher spin supercurrents for arbitrary spin massless supermultiplets and higher spin superfield cubic interactions},
	\href{https://doi.org/10.1007/JHEP08(2018)055}{JHEP \textbf{08} (2018), 055}
	[\href{https://arxiv.org/abs/1805.04413}{arXiv:1805.04413 [hep-th]}].
	
	\bibitem{Gates:2019cnl}
	S.~J.~Gates and K.~Koutrolikos,
	{\it Progress on cubic interactions of arbitrary superspin supermultiplets via gauge invariant supercurrents},
	\href{https://doi.org/10.1016/j.physletb.2019.134868}{Phys. Lett. B \textbf{797} (2019), 134868}
	[\href{https://arxiv.org/abs/1904.13336}{arXiv:1904.13336 [hep-th]}].
	
\bibitem{Zaigraev:2024ryg}
N.~Zaigraev,
{\it $\mathcal{N}=2$ higher-spin supercurrents},
\href{https://doi.org/10.1016/j.physletb.2024.139056}{Phys. Lett. B \textbf{858} (2024), 139056} [\href{https://arxiv.org/abs/2408.00668}{arXiv:2408.00668 [hep-th]}].

\bibitem{Zaigraev:2026jvu}
N.~Zaigraev,
{\it Structure of $\mathcal{N} = 2$ superfield higher-spin abelian cubic interactions},
[\href{https://arxiv.org/abs/2605.27206}{arXiv:2605.27206 [hep-th]}].

\bibitem{Buchbinder:2021ite}
I.~Buchbinder, E.~Ivanov and N.~Zaigraev,
{\it Unconstrained off-shell superfield formulation of 4D, $ \mathcal{N} $ = 2 supersymmetric higher spins},
\href{https://doi.org/10.1007/JHEP12(2021)016}{JHEP \textbf{12} (2021), 016}
[\href{https://arxiv.org/abs/2109.07639}{arXiv:2109.07639 [hep-th]}].

\bibitem{Buchbinder:2025yef}
E.~I.~Buchbinder, S.~M.~Kuzenko and I.~B.~Samsonov,
{\it Massless higher-spin supermultiplets in 5D harmonic superspace},
\href{https://doi.org/10.1007/JHEP02(2026)122}{JHEP \textbf{02} (2026), 122}
[\href{https://arxiv.org/abs/2509.06604}{arXiv:2509.06604 [hep-th]}].


\bibitem{Galperin:1984av}
A.~Galperin, E.~Ivanov, S.~Kalitzin, V.~Ogievetsky and E.~Sokatchev,
{\it Unconstrained $\mathcal{N}=2$ Matter, Yang-Mills and Supergravity Theories in Harmonic Superspace},
\href{https://doi.org/10.1088/0264-9381/1/5/004}{Class. Quant. Grav. \textbf{1} (1984), 469-498}
[erratum: \href{https://doi.org/10.1088/0264-9381/2/1/512}{Class. Quant. Grav. \textbf{2} (1985), 127}].





\bibitem{Galperin:1987ek}
A.~S.~Galperin, E.~A.~Ivanov, V.~I.~Ogievetsky and E.~Sokatchev,
{\it $\mathcal{N}=2$ Supergravity in Superspace: Different Versions and Matter Couplings},
\href{https://doi.org/10.1088/0264-9381/4/5/023}{Class. Quant. Grav. \textbf{4} (1987), 1255}.



\bibitem{Galperin:1987em}
A.~S.~Galperin, N.~A.~Ky and E.~Sokatchev,
{\it $\mathcal{N}=2$ Supergravity in Superspace: Solution to the Constraints},
\href{https://doi.org/10.1088/0264-9381/4/5/022}{Class. Quant. Grav. \textbf{4} (1987), 1235}.

\bibitem{Ivanov:2022vwc}
E.~Ivanov,
{\it $\mathcal{N}=2$ Supergravities in Harmonic Superspace}, in: Bambi, C., Modesto, L., Shapiro, I. (eds) \href{https://doi.org/10.1007/978-981-19-3079-9\_43-1}{Handbook of Quantum Gravity, Springer, 2024}
[\href{https://arxiv.org/abs/2212.07925}{arXiv:2212.07925 [hep-th]}].








\bibitem{Buchbinder:2025ceg}
I.~Buchbinder, E.~Ivanov and N.~Zaigraev,
{\it Towards $\mathcal{N}=2$ higher-spin supergravity},
[\href{https://arxiv.org/abs/2503.02438}{arXiv:2503.02438 [hep-th]}].

\bibitem{Fradkin:1985am}
E.~S.~Fradkin and A.~A.~Tseytlin,
{\it Conformal supergravity},
\href{https://doi.org/10.1016/0370-1573(85)90138-3}{Phys. Rept. \textbf{119} (1985), 233-362}.


\bibitem{Ivanov:2024gjo}
E.~Ivanov and N.~Zaigraev,
{\it Off-shell invariants of linearized $4D,\mathcal{N}=2$ supergravity in the harmonic approach},
\href{https://doi.org/10.1103/PhysRevD.110.066020}{Phys. Rev. D \textbf{110} (2024) no.6, 066020}
[\href{https://arxiv.org/abs/2407.08524}{arXiv:2407.08524 [hep-th]}].




\bibitem{Ivanov:2025mld}
E.~A.~Ivanov and N.~M.~Zaigraev,
{\it $\mathcal{N} = 2$ Supergravity and Harmonic Superspace: Linearized Supercurvatures},
\href{https://doi.org/10.1134/S106377962470148X}{Phys. Part. Nucl. \textbf{56} (2025) no.2, 247-252}.


\bibitem{Ivanov:2025gvs}
E.~Ivanov and N.~Zaigraev,
{\it Linearized $\mathcal{N}=2$ conformal supergravity in the harmonic approach},
[\href{https://arxiv.org/abs/2511.16325}{arXiv:2511.16325 [hep-th]}].



\end{thebibliography}
\end{document}